\documentclass[12pt]{article}
\usepackage{mathptmx}       
\usepackage{helvet}        
\usepackage{courier}        
\usepackage{type1cm}       
\usepackage[bottom]{footmisc}

\usepackage{url}

\usepackage{amsfonts}
\usepackage{amsmath}
\usepackage{amssymb}

\usepackage{enumerate}

\newcommand{\GNRR}{\mathbb R}
\newcommand{\GNNN}{\mathbb N}
\newtheorem{definition}{Definition}

\begin{document}

\begin{center}

\Large{\textbf{Transparent scatterers and transmission eigenvalues}}

\bigskip

\large{P.G. Grinevich\footnote{Steklov Mathematical Institute, RAS, Moscow, Russia;
    L.D. Landau Institute for Theoretical Physics, RAS, Moscow Region, Russia; Lomonosov Moscow State University, Department of Mechanics and Mathematics, Moscow, Russia; \textbf{e-mail:} \mbox{grinev@mi-ras.ru}} \  and R.G. Novikov\footnote{CMAP, CNRS, \'Ecole Polytechnique, Institut Polytechnique de Paris, France; \textbf{e-mail:} \mbox{roman.novikov@polytechnique.edu}}}

\end{center}

%\institute{P.G. Grinevich \at Steklov Mathematical Institute, RAS, Moscow, Russia;
%L.D. Landau Institute for Theoretical Physics, RAS, Moscow Region, Russia; Lomonosov Moscow State University, Department of Mechanics and Mathematics, Moscow, Russia
% \at \email{grinev@mi-ras.ru} % %order: first (given) name family name (first name abbreviated!)
% \and  R.G. Novikov \at CMAP, CNRS, \'Ecole Polytechnique, Institut Polytechnique de Paris, France
%\at \email{roman.novikov@polytechnique.edu}}  

\abstract{We give a short review of old and recent results on scatterers with transmission eigenvalues of infinite multiplicity, including transparent scatterers. Historically, these studies go back to the publications: Regge (Nuovo Cimento 14, 1959), Newton (J. Math. Phys. 3, 1962) and Sabatier (J. Math. Phys. 7, 1966). Our review is based on the works: Grinevich, Novikov (Commun. Math. Phys. 174, 1995; Eurasian Journal of Mathematical and Computer Applications 9(4), 2021; Russian Math. Surveys, 77(6), 2022). Results of the first of these works include examples of transparent at fixed energy potentials from the Schwartz class in two dimensions. The two others works include the result that, for compactly supported multipoint potentials of Bethe - Peierls - Thomas type in two and three dimensions, any positive energy is a transmission eigenvalue of infinite multiplicity.
} 

\medskip


%%%%%%%%%%%%%%%%%%%%
%%%%  Add main text here
%%%% Should containt at least three sections (e.g. Introduction, Main results, Conclusion)
%%%% Figures may be included as follows
% \begin{figure}
%	\centering
%	\includegraphics[width=0.7\linewidth]{}
%	\caption{}
%	\label{fig:}
%\end{figure}
%%%%%%%%%%%%%%%%%%%

\section{Introduction}
\label{sec:1}

To start with, we recall definition of scattering data for the stationary Schr\"odinger equation with a rapidly decaying at infinity potential, and we recall definition of boundary data for this equation in a bounded domain.

\subsection{Scattering data}
\label{sec:1.1}
We consider the stationary Schr\"odinger equation  
\begin{equation}
\label{GN_eq:1}  
-\Delta\psi + v(x)\psi = E\psi,  \ \ x\in{\GNRR}^d, \ \ d=1,2,3.
\end{equation}
We assume that potential $v(x)$ decays sufficiently fast as $|x|\rightarrow\infty$.

The scattering at potential $v(x)$ is described by the following solutions of equation~(\ref{GN_eq:1}):
\begin{equation}
\label{GN_eq:2} 
\psi^+= e^{ikx}+ f^+ \left(k,|k|\frac{x}{|x|}\right) \frac{e^{i|k||x|}}{|x|^{(d-1)/2}} +O\left(\frac{1}{|x|^{(d+1)/2}} \right), 
\end{equation}
as $|x|\rightarrow\infty$, for \textit{a priori} unknown $f^+=f^+(k,l)$ . Here $k,l\in\GNRR^d$, $k^2=l^2=E>0$.

The functions $f^+=f^+(k,l)$ is called  \textit{the scattering amplitude.}

It is also convenient to rewrite  $f^+(k, l)$ as:
\begin{align}
&f^+(k, l) = c(d, |k|)f(k, l), \ \ \mbox{where}\label{GN_eq:3}\\
&c(d, |k|) = -\pi i (-2\pi i)^{(d-1)/2}|k|^{(d-3)/2}, \, \text{ where } \sqrt{-2\pi i} = \sqrt{2\pi} e^{-i\pi/4}.\nonumber
\end{align}

\textit{The scattering operator} $\hat S=\hat S_E$ at a fixed energy level $E=\varkappa^2$, $\varkappa>0$, is defined by:
\begin{equation}
\label{GN_eq:4}  
(\hat S_E u)(\theta) =  u(\theta) -i\pi \varkappa^{d-2}  \int_{{\mathbb S}^{d-1}} f(\varkappa\theta', \varkappa\theta) u(\theta') d \theta',
\end{equation}
where $ u(\theta)$ is a test function, ${\mathbb S}^{d-1}$  is the unit sphere in  $\GNRR^d$, $\theta$,~$\theta'\in {\mathbb S}^{d-1}$, $d\theta'$ is the standard volume element at ${\mathbb S}^{d-1}$.

For more details see, for example, \cite{AGHH2005}, \cite{ChadanSabatier1989}, \cite{Faddeev1976}, \cite{Gelfand1957}, \cite{GRN1995}.

\subsection{Boundary data}

We also consider equation (\ref{GN_eq:1}) under the assumption that
\begin{equation}
\label{GN_eq:supp}
\mbox{supp} \ \ v \subset \cal D.
\end{equation}
where $\cal D$ is a connected bounded domain in $\GNRR^d$ with $C^2$ boundary, such that $\GNRR^d \backslash \overline{\cal D}$ is also connected. In this case we consider the Cauchy data ${\cal C}_v$  defined by:
\begin{equation}
\label{GN_eq:normal}
{\cal C}_v=\left\{\left(\left.\psi\vphantom{\frac{\partial \psi}{\partial\nu}}\right|_{\cal D},\left.\frac{\partial \psi}{\partial \vec n} \right|_{\cal D}\right)\ \ \ \parbox{6cm}{for all sufficiently smooth solutions $\psi$\\ of equation~(\ref{GN_eq:1}) in  $\bar{\cal D} ={\cal D}\cup\partial{\cal D}$ }  \right\},
\end{equation}
where $\frac{\partial \psi}{\partial \vec n}$ denotes the normal derivative, $\vec n$ is the outward normal vector to $\partial{\cal D}$.

These data can be also considered as the graph of an operator connecting the values of solutions $\psi$ and their normal derivatives at the boundary (Wigner operator in Gelfand's terminology \cite{Gelfand1957}). We consider the Cauchy data or the aforementioned operator as boundary data for equation~(\ref{GN_eq:1}) under assumption (\ref{GN_eq:supp}) . 

\subsection{Basic problems}

In this review we discuss examples of potentials in equation (\ref{GN_eq:1}), which are invisible from partial measurements of scattering or boundary data. In this connection in Section~\ref{sec:1.2} we discuss examples of transparent potentials in the sense of scattering amplitude vanishing at some energies, and in Sections~\ref{sec:1.3},~\ref{sec:1.4} we discuss potentials with transmission eigenvalues of finite and infinite multiplicity. In fact, transmission eigenvalues can be considered as eigenvalues of partial transparency. 

\section{Transparent potentials}
\label{sec:1.2}

\begin{definition}Potential $v=v(x)$ is called \textit{transparent at a fixed energy} $E>0$ if 
$$
\hat S_E\equiv \hat 1,
$$
where $\hat S_E$ is defined by (\ref{GN_eq:4}), that is
$$
f(k,l) = 0, \ \ \mbox{for all} \ \ k,l\in\GNRR^d \ \ \mbox{such that} \ \ k^2=l^2=E.
$$

\end{definition}

Historically, studies on transparent potentials in multidimensions go back to Tulio Regge \cite{Regge1959}, Roger Newton  \cite{Newton1962}, Pierre Sabatier \cite{Sabatier1966}. In particular, in these works, spherically symmetric ponentials transparent at a fixed positive energy $E$ were constructed for $d=3$. These potentials decay at infinity as $|x|^{-3/2}$. The Regge- Newton- Sabatier construction is based on a Gelfand-Levitan type equation for inverse scattering for spherically symmetric potentials at a fixed energy.

In turn, for $d=1$, the famous $N$-soliton potentials are reflectionsless for all positive energies. These potentials decay at infinity exponentially, see, for example, review paper \cite{Faddeev1976}. Let us make a simple but important observation: $N$-soliton potentials are transparent for $\left[\frac{N-1}{2}\right]$ energies, where $[\ ]$ denotes the integer part. These energies are given as $E_j=k_j^2$, where $k_j$ are positive real roots of the equation
$$
T(k)=1, \ \ \mbox{where} \ \  T(k)=\prod\limits_{j=1}^{N} \frac{k+i \varkappa_j}{k-i \varkappa_j}, \ \ \varkappa_j\in\GNRR, \ \  \varkappa_j>0.
$$
Here, $T(k)=s_{11}(k)=s_{22}(k)$ is the transmission coefficient for an $N$-soliton potential, where ${\cal E}_j=-\varkappa_j^2$ are the discrete energy levels for this potential; see \cite{Faddeev1976} for details.

More recently, in \cite{GRN1995} we constructed two-dimensional spherically symmetric real-valued potentials from the Schwartz class, which are transparent at a fixed positive energy $E$. The construction in \cite{GRN1995} essentially uses the two-dimensional inverse scattering transform at fixed energy developed in \cite{GrinevichManakov1986}. In turn, this construction uses the $\bar\partial$-approach to inverse scattering and also goes back historically to  Gelfand-Levitan type equations.

The problem of constructing three-dimensional non-zero transparent potentials with rapid decay at infinity is still open. On the other hand, there are no non-zero real regular transparent potentials with exponential decay at infinity in dimension $d\ge 2$; see \cite{Novikov1994} for $d\ge 3$ and \cite{GRN1995} for $d=2$.

Note that constructions of \cite{GrinevichManakov1986}, \cite{Novikov1994}, \cite{GRN1995} strongly use properties of Faddeev's growing solutions of the Schr\"odinger equation (\ref{GN_eq:1}) at fixed energy; see \cite{Faddeev1976}, \cite{GrinevichManakov1986}, \cite{Novikov1994}, \cite{GRN1995} for more details.   

Note also that the works \cite{Regge1959}, \cite{Newton1962}, \cite{Sabatier1966}, \cite{GRN1995}  were fulfilled before more recent studies on invisibility using cloaking!

Finally, recall that inverse scattering for the Schr\"odinger equation in dimension $d=1$ at all energies and in dimension $d=2$ at a fixed energy is deeply related with the soliton theory. This includes relations to the Korteweg -- de Vries equation in dimension $d=1$ (see \cite{Faddeev1976} and references therein), and to the S.~Novikov--Veselov equation in dimension $d=2$ (see \cite{NovikovVeselov1984}, \cite{GrinevichManakov1986}, \cite{GRN1995}).

\section{Transmission eigenvalues}
\label{sec:1.3}

\begin{definition}
\label{GN_def:2}  
Energy level $E$ is called \textit{transmission eigenvalue}, if the operator 
$$
\hat S_E-\hat 1
$$ 
has nontrivial kernel in $L^2({\mathbb S}^{d-1})$. Dimension of this kernel is called \textit{multiplicity of the transmission eigenvalue}. Here $\hat 1$ denotes the identity operator. 
\end{definition}

\begin{definition}
\label{GN_def:3}    
Energy level $E$ is called \textit{interior transmission eigenvalue} for equation (\ref{GN_eq:ite}) if there exists a pair of non-zero functions $\psi(x)$, $\phi(x)$ such that
\begin{equation}
\label{GN_eq:ite}  
-\Delta \psi(x) + v(x) \psi(x) = E  \psi(x), \ \ x\in{\cal D},
\end{equation}
$$
-\Delta \phi(x)  = E  \phi(x), \ \ x\in{\cal D},
$$
and
$$
\psi(x)\equiv \phi(x), \ \ \frac{\partial}{\partial\vec n}\psi(x)\equiv \frac{\partial}{\partial\vec n}\phi(x) \ \ \mbox{for all} \ \ x\in\partial{\cal D},
$$
where ${\cal D}$  and $\frac{\partial}{\partial\vec n}$  are as in  (\ref{GN_eq:supp}), (\ref{GN_eq:normal}).
\end{definition}

The transmission eigenvalue $E$ in Definition~\ref{GN_def:2}  can be treated as an energy of, at least, partial transparency in terms of the scattering operator $\hat S_E$, whereas the interior transmission eigenvalue $E$ in Definition~\ref{GN_def:3} can be treated as an energy of, at least, partial transparency in terms of the Cauchy data (\ref{GN_eq:normal}).

Historically, studies on transmission eigenvalues for scatterers with compact support go back to Kirsch \cite{Kirsch1986}, Colton and  Monk \cite{ColtonMonk1988}. In connection with more recent results in this direction, see, e.g. \cite{CakoniHaddar2012}, \cite{CakoniHaddar2013}, \cite{NguyenNguyen2017} and references therein.

A typical result of these studies is as follows:

\textit{For sufficiently regular compactly supported isotropic scatterers the transmission eigenvalues are discrete and have finite multiplicity.}

One of the purposes of this review is to attract the attention to the following facts.

\begin{itemize}
\item  This result is not valid for potentials from the Schwartz class, at least, in $\GNRR^2$. The point is that for transparent potentials constructed in \cite{GRN1995} and mentioned in Section~\ref{sec:1.2}, the energy of transparency $E$ is a transmission eigenvalue of infinite multiplicity. Moreover, in this case, the kernel of $\hat S_E-\hat 1$ coincides with the full space $L^2({\mathbb S}^{1})$.
\item  This result is not valid for multipoint scatteres of the  Bethe - Peierls - Thomas type in $\GNRR^d$, $d=2,3$,  discussed in Section~\ref{sec:1.4}, which are singular but have compact support. The point is that for these potentials, all positive energies are transmission eigenvalues of infinite mutiplicity. Moreover, in this case, the kernel of $\hat S_E-\hat 1$ has finite codimension  is $L^2({\mathbb S}^{d-1})$ for all $E>0$. In addition, for these potentials satisfying (\ref{GN_eq:supp}) any complex energy $E$ is an interior transmission eigenvalue of infinite multiplicity in ${\cal D}$.
\end{itemize}

\section{Multipoint scatterers}
\label{sec:1.4}

For $d=3$, the one-point scatterers in question were introduced by Bethe, Peierls (1935) and Thomas (1935) to describe the interaction between neutrons and protons. Subsequently, one-point and multipoint scatterers were considered by many authors, including Fermi (1936),  Zel'dovich (1960), Berezin and Faddeev~(1961). In particular, Fermi used such a model to explain strong interaction of slow neutrons with nuclei. For references and more details, see, for example, monograph \cite{AGHH2005}. Possible generalizations are discussed in \cite{Kurasov2009}, \cite{Chashchin2018}.

The most intuitive definition of point scatterers is as follows. Consider a sequence of regular compactly supported potentials 
$$
v_N(x), \ \ N\in\GNNN,
$$
such that for $N\rightarrow\infty$:
\begin{itemize}
\item Diameter of $\mbox{supp}\ v_N(x)$ converges to 0;
\item at any compact interval of energies the scattering amplitude $f_N^+(k,l)$ converges to well-defined non-trivial limit $f_{\infty}^+(k,l)$.
\end{itemize}

For $d=1$, these scatterers are standard Dirac $\delta$-functions with some coefficients.  For $d=2,3$, the simplest scatterers of such a type are the Bethe-Peierls-Thomas-Fermi scatterers; in this case the sequence $v_N(x)$ converges to 0 in the sense of distributions. 

To construct these scatterers, one can start with
$$
v_N(x) = \phi(N) v_1(N x),
$$
$v_1(r)$ is the characteristic function of unit ball in $\GNRR^d$, $d=1,2,3$. To obtain a good limit for $f_{\infty}^+(k,l)$, one has to assume, in particular, that
$$
\phi(N) \sim \left\{\begin{array}{ll} N, & d=1,\\  N^2/\log(N), &  d=2,\\  N^2 &  d=3,
\end{array}  \right.
$$
up to coefficients and lower terms.

By specifying properly the behavior of $\phi(N)$, we obtain, for each $d=1,2,3$, a family of Dirac-Bethe-Peierls-Thomas-Fermi scatterers $\delta_{\alpha}(x)$, supported at $x=0$ and parameterized by $\alpha\in\GNRR\cup\infty$. The number $1/\alpha$ can be interpreted as the \textit{strength of the scatterer}. Sometimes, for $d=2,3$, these $\delta_{\alpha}(x)$ are called ``renormalized $\delta$-functions''. For $d\ge4$, the Green function for the Helmholtz operator $\Delta +E$ does not belong locally to $L^2(\GNRR^{d})$, and, as a corollary, the aforementioned construction of point potentials does not work.

In addition, $\psi$ satisfies the stationary Schr\"odiger equation
$$
-\Delta\psi + v(x)\psi = E\psi,
$$
with $n$-point ponential
$$
v(x)=\sum\limits_{j=1}^n \delta_{\alpha_j}(x-y_j)
$$
iff
$$
-\Delta\psi(x) = E\psi(x),  \ \ \mbox{for all}\ \ x\ne y_j, \ \ j=1,\ldots,n,
$$
and near the points $y_1$,\ldots,$y_n$:
\begin{enumerate}
\item If $d=1$, then $\psi(x)$ is continuous at $x=y_j$, and its first derivative has a jump
$$ 
-\alpha_j\left[\psi'(y_j+0)- \psi'(y_j-0)\right] = \psi(y_j);
$$
\item If $d=2$, then
\begin{eqnarray*}
  &\hphantom{and}& \ \ \ \psi(x) = \psi_{j,-1}\ln|x-y_j| + \psi_{j,0} + O(|x-y_j|) \ \ \mbox{as} \ \ x\rightarrow y_j, \\
  &\mbox{and}& \ \ \ \left[-2 \pi \alpha_j -\ln 2 + \gamma\right] \psi_{j,-1} = \psi_{j,0},
\end{eqnarray*}
where $\gamma=0.577\ldots$ is the Euler's constant.
\item If $d=3$, then
\begin{eqnarray*}
 &\hphantom{and}& \ \ \ \psi(x) = \frac{\psi_{j,-1}}{|x-y_j|}+ \psi_{j,0} + O(|x-y_j|)\ \ \mbox{as} \ \ x\rightarrow y_j, \\
 &\mbox{and}& \ \ \ 4\pi \alpha_j \psi_{j,-1} = \psi_{j,0}.
\end{eqnarray*}
\end{enumerate}

The Schr\"odinger equation (\ref{GN_eq:1}) with multipoint potentials as above is exactly solvable! In particular, it is known that 
$$
\psi^+(x,k)=e^{ikx}  +  \sum\limits_{j=1}^n q_j(k) G^+(x-y_j,|k|^2),
$$
\begin{equation}
\label{GN_eq:5}  
f(k,l) = \frac{1}{(2\pi)^d}  \sum\limits_{j=1}^n q_{j}(k) e^{-il y_j},
\end{equation}
where $G^+(x, E)$ denotes the Green functions for the Helmholtz operator with the Sommerfeld radiation condition, see, for example, \cite{GRN2021}.

The vector of constants $q(k)=(q_{1}(k),\ldots,q_{n}(k))^t$ is defined as the solution of the linear system:
$$
A(|k|) q(k) = b(k),
$$
with the $n\times n$ matrix $A$ and the right-hand side vector $b(k)$ defined by:
\begin{align*}
&A_{j,j}(|k|) &= &\ \ \alpha_j - i(4\pi)^{-1}|k| , \ \ &d=3,\nonumber\\ 
\label{GN_eq:2.11}
  &A_{j,j}(|k|) &= &\ \ \alpha_j - (4\pi)^{-1}(\pi i -2 \ln(|k|)), \ \ &d=2,\\
 &A_{j,j}(|k|) &= &\ \ \alpha_j + (2 i |k|)^{-1}, \ \ &d=1,\nonumber\\  
  &A_{j,j'}(|k|) &= &\ \ G^{+}(y_j-y_{j'},|k|^2), \ \ &j'\ne j,\nonumber\\
  &b_{j}(k) &= &-e^{iky_j}.
\end{align*}

Formula~(\ref{GN_eq:5}) for $f(k,l)$ implies that each positive energy $E$ for the aforementioned multipoint potentials is a transmission eigenvalue of infinity multiplicity, since the operator $\hat S_E- \hat 1$ has rank at most $n$. For more details on transmission and interior transmission eigenvalues for the  multipoint potentials, see \cite{GRN2021}, \cite{GRN2022}.

In particular, in \cite{GRN2022} we continued studies on multipoint scatterers by considering Schr\"odinger's equation with potential which is a sum of a regular function and a finite number of point scatterers of Bethe-Peierls-Thomas type. For this equation we considered the spectral problem with homogeneous linear boundary conditions, which covers the Dirichlet, Neumann, and Robin cases.

In \cite{GRN2022}, we showed that if the energy $E$ is an eigenvalue with multiplicity $m$ for the regular potential, it remains an eigenvalue with multiplicity at least $m-n$ after adding $n < m$ point scatterers.

As a consequence, because for the zero potential all energies are transmission eigenvalues with infinite multiplicity, this property also holds for $n$-point scatterers as we mentioned above.

More recently, we also observed that a converse inequality also holds for the aforementioned boundary conditions in \cite{GRN2022}.  We found that if the energy $E$ is an eigenvalue with multiplicity $m$ for a sum of a regular potential and an $n$-point potential, $n < m$, then this energy is an eigenvalue with multiplicity at least $m-n$ for the regular potential.

Note also that there is no non-zero real-valued transparent at a fixed positive energy  multipoint potentials of Bethe-Peierls-Thomas type is dimensions $d=2,3$, see \cite{KuoNovikov2024}.

 %%%%%%%%%%%%%%%%%%%%%%%%%%%%%%

\end{document}